# Data-Driven and SE-assisted AI Model Signal-Awareness Enhancement and Introspection

Sahil Suneja
IBM Research
Yorktown Heights, NY, USA
suneja@us.ibm.com

Yufan Zhuang
IBM Research
Yorktown Heights, NY, USA
yufan.zhuang@ibm.com

Yunhui Zheng
IBM Research
Yorktown Heights, NY, USA
zhengyu@us.ibm.com

Jim A. Laredo
IBM Research
Yorktown Heights, NY, USA
laredoj@us.ibm.com

Alessandro Morari
IBM Research
Yorktown Heights, NY, USA
amorari@us.ibm.com

## ABSTRACT

AI modeling for source code understanding tasks has been making significant progress, and is being adopted in production development pipelines. However, reliability concerns, especially whether the models are actually learning task-related aspects of source code, are being raised. While recent model-probing approaches have observed a lack of signal awareness in many AI-for-code models, i.e. models not capturing task-relevant signals, they do not offer solutions to rectify this problem. In this paper, we explore data-driven approaches to enhance models' signal-awareness: 1) we combine the SE concept of code complexity with the AI technique of curriculum learning; 2) we incorporate SE assistance into AI models by customizing Delta Debugging to generate simplified signal-preserving programs, augmenting them to the training dataset. With our techniques, we achieve up to 4.8x improvement in model signal awareness. Using the notion of code complexity, we further present a novel model learning introspection approach from the perspective of the dataset.

## 1 INTRODUCTION

Over the past few years, AI models have made significant progress in source code understanding tasks such as defect detection [20, 24, 26, 39, 43, 60], code summarization [7, 29, 33, 53], code completion [18, 40], bug fixing [69, 70], function and variable naming [4, 5, 10, 41], code recommendation [34, 51], amongst others [12, 75]. The wide availability of open source codebases has fueled this progress, and we are starting to see the adoption of such AI-for-code models in software development workflows [8, 25, 68]. Accompanying their rapid proliferation is an emerging scrutiny over the models' quality. This goes beyond just the usual concerns about the black-box nature of AI modeling inhibiting a direct deduction of its learned logic. Reliability concerns are being raised along multiple dimensions including data duplication bias, labeling quality, low robustness and generalization, amongst others [3, 9, 16, 46, 54, 81].

Recently, concerns have been raised regarding whether the models are actually learning task-relevant aspects of source code, especially when small input perturbations may lead to incorrect predictions [13, 57, 67]. For example, in a vulnerability detection setting, a model is free to select any differentiating code features in its endeavour to learn a separator between buggy and healthy code samples. It can arrive at this separator by using non-representative signals to the task at hand, such as unexpected correlations between code samples, or certain programming constructs, or identifier names, which may happen to differ for buggy or healthy samples. Learning such a separator may even yield great classification performance, which is perfectly valid from a model's perspective in a theoretical setting. However, in a practical setting, it is a risky proposition to employ models influenced heavily by such dataset nuances, as opposed to capturing task-relevant signals. Today's AI-for-code models have been shown to significantly lack such 'signal awareness' [57, 67], where the great F1 and accuracy values seem to be achieved by learning code features not relevant to the task at hand.

While existing model-probing approaches expose the lack of signal awareness in AI-for-code models, they do not offer solutions to rectify this aspect of the model's quality. Advancements in AI research in general, lead us to believe in the models' potential to pick such relevant signals with appropriate *guidance* during training, which is one of the themes we explore in this paper. Furthermore, despite the initial attempts of these approaches to uncover black-box model learning, unresolved gaps still remain. While such approaches can detect *if* the models are learning task-relevant signals, they do not identify *what* aspects of source code are the models learning. This can be a valuable resource towards uncovering the logic learned by models, and assert trust in them as they integrate with the software development workflows.

Along the lines of AI-for-code model-quality research, in this paper we explore model assistance and introspection mechanisms. We first present two approaches to enhance signal awareness, which are data-driven and use artifacts specific to source code, related to its complexity. Then, continuing along the data perspective, we present a code-complexity-driven approach to deduce model learning behavior. Figure 1 presents a view of all our approaches co-existing within the general model training pipeline.

Our first approach to enhance model signal awareness marries the SE concept of code complexity, with the AI technique of curriculum learning [11]. Specifically, we incorporate the notion of code complexity metrics by feeding program samples to the model in increasing order of their code complexity. The intuition is that by presenting 'easier' examples first, it would improve the model's chances of picking up task-relevant signals, helping it to sift-through noise with more complex examples down the line.

Our second approach to improve model signal awareness is via program-simplification-based data augmentation. As with the previous approach, we incorporate SE assistance into AI model training by customizing Delta Debugging [79] to generate simplified program samples. The intuition is that by adding smaller and potentially de-noised code samples to the training dataset, while preserving their task profile, the model can potentially learn relevant signals better. In context of vulnerability detection task, compared to existing source-level bug-seeding-based augmentation methods which may produce unseen bugs [14, 22, 48, 55, 56, 59, 71], generating simpler programs while preserving the existing bugs is the key difference in our approach, enabling it to assist models in better understanding task-relevant code constructs.

Next, we continue our data-driven model exploration beyond the model learning enhancement approaches. We leverage the unique opportunity afforded by the source code setting and use SE techniques for black-box model learning introspection. Specifically, we present a code complexity driven approach to analyze a model's predictions from the dataset perspective, beyond the statistical measures of the model prediction accuracy. The intuition is to analyze the characteristics of the samples which the model predicted correctly versus those that it got wrong. By using code complexity metrics to group samples by prediction accuracy, our approach uncovers model learning behavior in terms of the aspects of code the model can grasp, and where it faces difficulties. This has versatile applications such as deciphering learned model-logic, driving model evolution, model design space evaluation, and dataset segmentation, amongst others.

Being data-driven, our approaches requires no knowledge about the target model's internals, and are applicable to all model types. Additionally, they are task-agnostic and programming-language-agnostic, broadening their applicability further. We apply these on three different neural network architectures operating on different representations of source code- convolutional neural network (CNN), recurrent neural network (RNN), and graph neural network (GNN). Results show substantial improvements in the models' signal awareness across different datasets, when assisted with our proposed approaches. Complexity-ranked training can provide up to 32% signal awareness boost, whereas program simplification based augmentation surpasses it, realizing up to 4.8x improvement. The higher improvements achieved by the latter justifies it's somewhat complex requirements (e.g. existence of a sample labeler), when compared to relatively straightforward code complexity extraction and ranking. This highlights a trade-off between the resources available to deploy towards model learning enhancement, versus the level of signal awareness improvements expected.

To deduce what's happening under the covers, we use our code-complexity-driven model introspection approach to trace model evolution across augmentation iterations. Amongst the model learning insights provided by our approach, it is the issue of the models facing difficulty in understanding bigger (and more complex) code samples. Beyond providing support for this perhaps intuitive expectation, a more intriguing result is about augmentation helping the models learn more complex code. This observation is especially unique since the augmentation scheme employed, like the model training in this case, was agnostic of code complexity.

Existing AI-for-code model explanation approaches, while being scarce, tend to use white-box model internals to add some transparency into the model logic. This typically involves using the models' trained variables, such as gradients or attention weights, to determine the importance of different features/regions of the input data/code [21, 60]. The approximate nature of such mappings, from the model side back to the data, can make them misleading [2]. This approximation is also seen in some black-box explanation approaches [57] which do not require the inputs to remain natural or valid. Furthermore, unlike our work, these do not tie source code constructs to model signal awareness. Instead, we approach the problem from the opposite direction- from the dataset's perspective, based on concretely defined characteristics of source code. This empowers our model learning introspection approach to offer more code-centric and developer-friendly insights than the usual quantitative measures of model performance, as well as existing model explanation approaches.

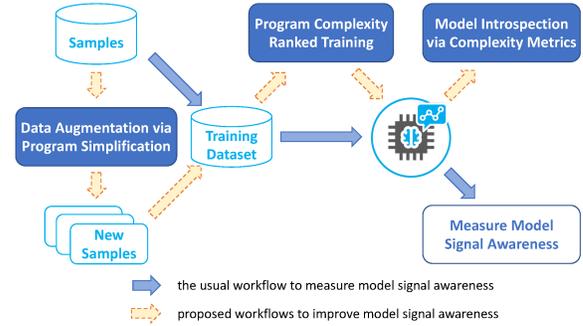

Figure 1: Model signal-awareness enhancement and introspection, augmenting the usual model training and measurement workflow.

With this work, we attempt to assist AI-for-code in better utilizing its potential, guiding it towards task-relevant learning, and uncovering actionable insights for targeted model advancement. By utilizing proven SE techniques to help assert confidence in the model's reliability, our hope is to highlight the potential of realizing better quality AI to assist with software development. The main contributions of our work are as follows:

- We present what we believe to be the first attempt at improving signal awareness of AI-for-code models, using SE concepts to assist AI understanding of source code.
- We tailor AI's curriculum learning for the source code domain, by coupling it with the notion of code complexity.
- We show the superiority of targeted augmentation with simplified programs, over generic augmentation, to improve model learning.
- We present a novel perspective for deducing model learning behavior using code complexity of the dataset.

## 2 MOTIVATION

Recent work has shown that even though AI-for-code models fare well on the usual statistical measures model quality (F1, accuracy, etc.), their signal awareness is low [67]. That is, the models aren't necessarily picking up task-relevant signals while learning to differentiate samples belonging to separate classes (e.g. 'buggy' or 'healthy' code). This can happen when the models pick up non-representative code features to the task at hand, such as unexpected correlations between samples and certain keywords, or programming constructs like ifs / loops / variable names, which may be more

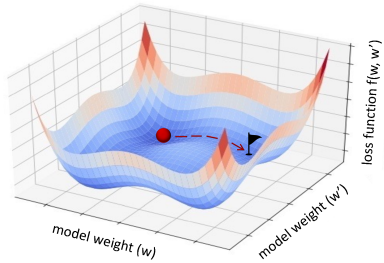

Figure 2: Intuition: assist model to reach alternate minima (the flag), maintaining similar performance as achieved in original minima (the ball), but potentially capturing more task-relevant features.

```
void bad()
{
  wchar_t * data = NULL;
  data = new wchar_t[10];
  {
    wchar_t source[10+1] = SRC_STRING;
    size_t i, sourceLen;
    sourceLen = wcslen(source);
    for (i = 0; i < sourceLen + 1; i++)
    {
      data[i] = source[i];
    }
    printWLine(data);
    delete [] data;
  }
}
```
(a) Original code
Model Prediction: Buggy
Ground Truth: Buggy

```
void bad()
{
  wchar_t *data = NULL;
  data = new wchar_t[10];
  {
    wchar_t source[10+1] = SRC_STRING;

    size_t i = wcslen(source);
    for (;;)

         (data);

  }
}
```
(b) 1-Minimal
Model Prediction: Buggy
Ground Truth: **Non-buggy**

Figure 3: Example showing lack of signal awareness in an AI-for-code model (F1: 97) for vulnerability detection. Comparing original code vs. it's extracted 1-minimal, but incorrectly predicted by model as buggy even when it doesn't contain the actual bug (highlighted in red). This suggests model learns signals not relevant to task [67].

prevalent in one sample class than the other. Learning class separators in this way is perfectly valid from the model's perspective, but such a model isn't necessarily 'understanding' task-relevant aspects of source code correctly, but merely learning dataset nuances. Such a model is prone to failures when applied in real-world settings, contributing to model reliability concerns [1, 9].

We believe in the potential of the models to overcome this 'weakness' of lacking signal-awareness with appropriate 'guidance' during training, and seek SE concepts to provide this guidance in a code-specific fashion. The model itself is doing its job well, learning how best to separate samples based on available input features, to reach a local minima in the loss landscape. But, as shown in Figure 2, there can exist other local minima in this landscape, which can offer similar model performance, but which rely on features more in sync with the task expectation. If we can somehow nudge the model to reach an alternate minima, while using domain-specific assistance, perhaps the model's signal awareness can be improved.

These sort of issues exist in AI in general, and are typically solved by white-box or domain-specific approaches such as robust training and adversarial training [27, 45, 80]. In this work, we explore code-centric, data-driven approaches to assist AI-for-code models in focusing more on task-relevant aspects of source code. We do this by leveraging SE techniques to incorporate notion of complexity of code into model learning, while treating the model as a black-box. In addition to improving model signal awareness, we additionally probe this model black-box, again from the perspective of the dataset's code complexity, to reveal code-centric developer-friendly insights to further assist with model learning.

## 3 BACKGROUND

We briefly cover Delta Debugging (DD), which we use to simplify program samples for our augmentation technique, and describe how it has also been employed for probing AI model signal awareness.

**Delta Debugging** [79] was first proposed to minimize long crash inducing bug reports for Mozilla's web browser. The input to DD is a sequence satisfying some predefined oracle. The goal is to find a subset of the input satisfying the following two requirements: (1) the subset leads to the same outcome; and (2) not a single element can be removed to preserve the outcome. Such a subset is called *1-minimal*. Please refer [79] for details. Our intuition behind program simplification comes from this failure-inducing input simplification idea. We translate DD's process of continuously reducing the input while maintaining the same output, into successively simplifying the input program while maintaining its task profile (e.g. vulnerability profile in case of a vulnerability detection setting). Each valid iteration generates code simpler than the parent, which we augment to the original dataset for subsequent model training.

**Model Signal Awareness:** Recent works [57, 67] have used DD to probe AI-for-code models for explaining them, or exposing a lack of signal awareness in them. In the aforementioned setting, the Mozilla target is replaced with a prediction model, and the failure-inducing HTML file with a program sample predicted to be of a particular class by the model under test. The DD cycle is then employed to identify the minimal sub-program (1-minimal) which preserves the model's prediction. Finally, the model's signal-awareness is determined by testing if the 1-minimal exhibits the same task profile as the original sample (e.g. containing the same bug as the original sample). Widespread occurrence of cases as shown in Figure 3 expose the lack of signal awareness in today's AI-for-code models, despite their high F1 / accuracy scores.

## 4 DESIGN

Figure 1 presents a high-level overview of all our data-driven approaches co-existing within the typical model AI modeling pipeline. Our program-simplification-based augmentation technique resides in the dataset curation phase, while our complexity-ranking approach is applicable during the model training phase. Finally, once a trained model is available, our introspection approach comes into play to deduce model learning behavior from the dataset's perspective. All techniques can exist independently of each other.

### 4.1 Code-complexity-ranked Training

This model training approach marries the AI concept of curriculum learning [11] with the SE notion of source code complexity. The idea is to present the model less-complex code samples initially during training, and to increase sample complexity progressively, mimicking the human learning procedure. Coming from the original training set, these initial samples still contain the same traits which we want the model to learn, while being relatively 'easier' than their counterparts. This can improve the model's chances of picking task-relevant signals better, with less interference from potential 'noise' existing in more-complex samples- in the form of statements or constructs not relevant to the task at hand. The hope is that equipped with the knowledge of the right 'signals' to look for, the model may be better able to sift through noise in the rest of the samples, and refine its learning while maintaining task-awareness.

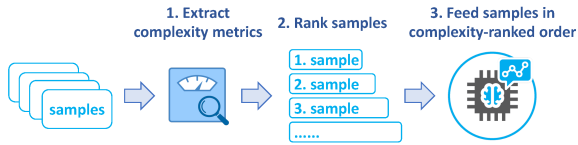

Figure 4: Our complexity-ranked training approach: 1. Extract code complexity of training samples. 2. Rank them in increasing order of complexity. 3. Feed them to model in complexity-ranked order.

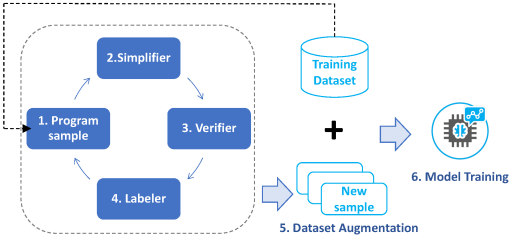

Figure 5: Our program-simplification-based augmentation approach: 1-2. For a training set sample, reduce it via DD. 3-4: Validate, label and emit reduced sample; continue reduction. 5-6: Train model on training set augmented with all simplified samples.

Figure 4 presents the steps involved in our complexity-ranking training approach. First, the code complexity metrics are extracted for training samples. There exist multiple such metrics ranging from straightforward lines of code (`sloc`), to higher-order `cyclomatic` and `halstead` complexity metrics, measuring concepts such as linearly independent paths, coding effort required, amongst others. Next, the training set samples are ranked in the increasing order of their corresponding metric(s) score. Finally, the samples are then fed to the model for training in their complexity-ranked order (e.g. `difficulty 12 − > 17` in left half of Figure 7), as opposed to generic random-sampling based training. As we show in Section 6.1, introducing code complexity awareness into model training in this straightforward manner, can improve the model's signal awareness significantly. The multiple complexity metrics options, in this setting, serve as tunable knobs to influence model learning.

### 4.2 Augmentation via Program Simplification

Here, we take a different route to assist model learning, with the notion of complexity of code being implicit, in contrast to the previous approach. What remains same is the data-driven methodology, as well as the SE-assistance to AI model training. This time around, we use the data augmentation route. However, instead of just offering more examples, we "simplify" them while preserving the signals.

Figure 5 presents the overall flow. We borrow the popular fault-isolation technique of Delta Debugging [79] (*Simplifier* in the figure) to generate simplified program samples from the training dataset. We follow the same procedure as outlined in Section 3, wherein Delta Debugging (DD) reduces the program samples at the level of source code tokens. For each input sample, the intermediate subprograms generated during the reduction cycle, which satisfy certain prerequisites, are emitted to serve as augmentation candidates. With each valid iteration of the reduction cycle generating code smaller than the parent, the successive denoising achieved as a result carries the potential of assisting the model in learning the relevant signals better, when trained on the augmented dataset. Figure 7 shows the effect of our augmentation approach adding simpler examples, in terms of the dataset complexity distribution.

| # | Program | Valid | Vul |
|---|---------|-------|-----|
| 1 | `void foo (int a) {int b = 10; int buf[10]; a + 3; buf[b] = 1;}` | ✓ | ✓ |
| 2 | `                                         10]; a + 3; buf[b] = 1;}` | ✗ | |
| 3 | `void foo (int a) {int b = 10; int buf[` | ✗ | |
| 4 | `                   int b = 10; int buf[10]; a + 3; buf[b] = 1;}` | ✗ | |
| 5 | `void foo (int a) {          [10]; a + 3; buf[b] = 1;}` | ✗ | |
| ... | ...... | | |
| 13 | `void foo (int a) {int b = 10; int buf[10];          buf[b] = 1;}` | ✓ | ✓ |
| ... | ...... | | |
| 21 | `void foo (int a) {int b = 10; int buf[10];          buf[   = 1;}` | ✗ | |
| 22 | `void foo (int a) {int b = 10; int buf[10];          buf[   ;}` | ✓ | ✓ |
| ... | ...... | | |

Figure 6: Generate signal-preserving simplified programs. Iteration #1 shows a valid but vulnerable program. In #2-3, DD first cuts it into half but fails to find a valid sub-program. DD repeatedly tries a finer granularity until reaching the token level. It finds two valid and vulnerable subprograms in #13 and #22.

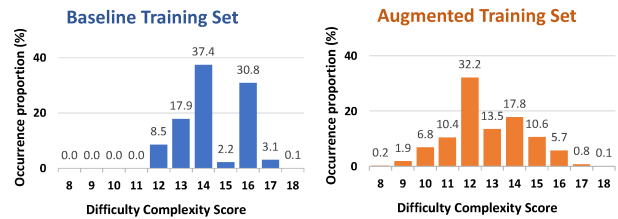

Figure 7: Complexity distribution of s-bAbI training set before and after augmentation.

The iterative DD process is driven by an oracle, which decides whether or not an intermediate reduced subprogram should be picked for subsequent reductions. We customize the oracle with a *Verifier* and a *Labeler* to require the reduced subprograms to satisfy the following properties:

- *Valid program (Verifier)*. We enforce that the reduced subprogram is valid and compilable. Although we may be able to reduce more tokens by dropping this requirement, 'simplifying' the parent sample even more, this would be counter-productive to later train models on incorrect code.
- *Vulnerability type (Labeler)*. We additionally check the reduced subprogram for either possessing the same bug as the original sample, or being bug-free. By ensuring no new bug gets introduced as part of the reduction cycle, this maintains the dataset integrity, so as not to emit samples with out-of-dataset labels.

Each reduced subprogram satisfying above properties is correspondingly labeled and emitted, and the reduction cycle continues. Figure 6 shows a few key steps to illustrate this process.

The overall reduction cycle results in multiple simplified samples being generated from each training set sample, with each valid iteration generating code smaller than the parent. The final step is then to assist model learning by adding these smaller, potentially de-noised code samples into the training mix.

While we use the program simplification aspect to improve model signal awareness, our approach can be used as a standalone augmentation scheme to either reduce overfitting- by adding all generated samples to the training set, or to reduce class imbalance- by adding only minority-class generated samples. It can also be combined with our complexity-ranked training approach, by (i) applying the latter atop the augmented set to assist model learning, or (ii) training and comparing the model on subsets ordered by program simplicity to verify model capacity and quality.

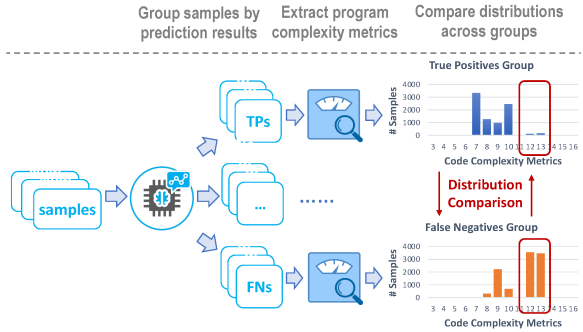

**Figure 8: Model Learning Introspection via Complexity Metrics.**

*4.2.1 Discussion.* **1. Labeler Customization.** While we presented our approach in the setting of vulnerability detection, it is independent of the target source code understanding task. The labeler is what tailors our approach to a particular task, and controls the quality of the generated samples. Some options, highlighting a cost vs. quality trade-off, include: human domain expert, original dataset labeler, line-based code-feature matcher, static analyzer, and fuzzer, amongst others. For our experiment settings and the datasets considered in this paper (Section 5.1), we found the Infer analyzer [23] to work quite well. Specifically, at each iteration of the reduction cycle, we compare Infer's analysis of the reduced subprogram with that of the original program sample, to ensure that the reduced subprogram is either bug-free, or possesses only the same bug as the original program sample. The latter is detected by a hit for the original bug in Infer's `preexisting.json` and a miss in `introduced.json` analysis comparison files. Although Infer as a labeler is a fortunate fit given our target datasets, as discussed above, our approach is not reliant on it.

**2. Training distribution.** During augmentation, we are adding additional data to the training set, which can be considered as sampling around the original data points[17], but in the discrete program space and with an accurate labeler. Theoretically, this may add learning difficulties for models due to the further complicated decision boundary as the data distribution shifts away from the test set. But more importantly, it would also help the model to learn real signals, therefore improving its reliability. In practice, we observed that the model performance did not deteriorate with our simplified-program augmentation, while its signal awareness increased dramatically (Appendix Table A.1).

## 4.3 Model Learning Introspection

Moving beyond the model learning enhancement, but while still maintaining the notion of code complexity, we tackle another awareness-related aspect of AI-for-code models- uncovering the black-box learning. Unlike the few existing model explanation approaches, we approach model learning analysis from the dataset's perspective, using well-defined source code characteristics to deduce model learning. The intuition is to compare the common characteristics of the samples which the model predicted correctly vs. those that it got wrong, and using it to highlight the aspects of code the model is able to grasp, vs. those where it's facing difficulties.

Figure 8 shows the overall flow of our approach. Given a trained model and the corresponding model predictions for the test set samples, we first extract their complexity metrics as discussed in Section 4.1. Then, we group the samples by their prediction accuracy, e.g. True Positives and False Negatives. For each group, we generate a distribution of the samples with respect to their scores for a given complexity metric. Finally, we compare and contrast the complexity metrics distributions across groups, to concretely highlight the differing aspects of code grasped or missed by the model, driven by the particular complexity metric employed for analysis. Figure 8 shows an example where the model's suffering from a lot of mispredictions on code samples with high complexity (generic, for exemplification), as indicated by the highlighted FN bars having negligible presence in their TP distribution counterparts.

This SE-assisted, data-driven model prediction analysis approach can be used to drive a multitude of introspection use-cases, such as:

(1) Dataset segmentation & introspection: Which samples are easy vs. hard for prediction? What are the characteristics that make samples easy or hard to predict? Are there any commonalities, bias or limitation in the dataset affecting different models?
(2) Decipher learned model-logic: What aspects of code is the model able to grasp, and where is it facing difficulties?
(3) Derive model signal awareness: Are the signals learned from code samples relevant to the task, or just noise / dataset bias?
(4) Drive code-centric model evolution: Target code characteristics common to mispredictions in training. Trace model understanding improvements across hyperparameter tuning iterations.
(5) Design space evaluation from dataset perspective: Trace model improvement techniques such as data augmentation, curriculum learning, active learning, adversarial training, etc.
(6) Code-centric model comparison: Assess task-suitability across models, especially useful across models with similar performance measures (F1, accuracy, etc.)

In our experiments, we cover two of these use-cases (2 and 5)- figuring out how model learning evolves, in the context of model signal awareness, across the iterations of our aforementioned program-simplification-based augmentation.

## 5 EXPERIMENT CONFIGURATION

We now describe the setup we use, in terms of the datasets, models and metrics for evaluating our approaches.

### 5.1 Datasets

We use a vulnerability detection use-case for evaluating our approaches. We use the datasets employed by [67], which are amenable to measuring the signal awareness of models trained upon them, and thus its enhancement by our approaches.

**s-bAbI:** The s-bAbI synthetic dataset [62] contains syntactically-valid C programs with non-trivial control flow, focusing solely on the buffer overflow vulnerability. We used the s-bAbI generator to create a balanced dataset of almost 40K training functions. Samples with 'UNSAFE' tag are labeled as 1, and those with 'SAFE' tag as 0.

**Juliet:** The Juliet Test Suite [52] contains synthetic examples with different vulnerability types, designed for testing static analyzers. From its test cases, we extract almost 32K training functions, amongst which 30% are vulnerable. Samples tagged as 'bad', and with clear bug information as per Juliet's `manifest.xml` file, are labeled as 1, while the ones with a 'good' tag are labeled as 0.

**D2A:** D2A [82] is a real-world vulnerability detection dataset built over multiple Github projects- `OpenSSL`, `FFMpeg`, `HTTPD`, `Nginx` and `libtiff`. It contains in-depth trace-level bug information, derived using differential analysis atop the Infer static analyzer outputs of consecutive repository versions, before and after bug-fixing commits. Function-level sample extraction from D2A traces yields 6728 functions, as done by [67].

## 5.2 Models

We apply our learning enhancement approaches to three popular neural network architectures for vulnerability detection tasks.

A **Convolutional Neural Network (CNN)** tries to learn the pictorial relationship between tokens and underlying bugs. Similar to [60], we normalize the function names and variable names to fixed tokens such as `Func` and `Var`. We set the embedding layer dimension to 13, followed by a 2d-convolutional layer with input channel as 1, output channel as 512, and kernel size as (9, 13). The final prediction is generated by a 3-layer multilayer perceptron (MLP) with output dimensions being 64, 16, and 2.

A **Recurrent Neural Network (RNN)** treats a program as a linear sequence of tokens to learn the temporal relationship between the tokens and bugs. We implement our RNN based on [38]. We set the embedding layer dimension as 500, followed by a two-layer bi-directional GRU module with hidden size equals to 256. The final prediction is generated by a single-layer MLP. Similar to CNN, we normalize the input functions as well.

A **Graph Neural Network (GNN)** operates on the graph-level representations of source code, which are commonly used in program analysis and compilation. For example, in [66, 84], GNN tries to learn bug patterns in a Code Property Graph [74]. We set the embedding size as 64, followed by a GGNN layer [37] with hidden size 256 and 5 unrolling time steps. Similar to [84], we do not normalize the tokens. The node representations are obtained via summation of all node tokens' embedding, and the graph representation read-out is constructed as a global attention layer. The final prediction is generated by a 2-layer MLP with output dimensions 256 and 2.

The models are trained over the datasets presented in Section 5.1, using a 80:10:10 train:validate:test split. For all models, we set dropout rate as 0.2 during training, and used the Adam [35] optimizer. We tuned learning rate in $\{10^{-3}, 10^{-4}\}$ and batch size in $\{24, 36, 128, 256, 512\}$. Models are trained to minimize cross entropy loss. We save the checkpoint with the least validation loss across epochs, with early stopping employed (patience = 10).

## 5.3 Tools and Metrics

Here we present the methodology we adopt for code complexity extraction and signal-awareness measurement.

**Code Complexity Metrics Extraction:** We used Frama-C [19] to extract complexity metrics for s-bAbI and Juliet samples. In addition to the relatively straightforward counters such as `sloc`, `ifs`, `loops`, we also collected higher-order `cyclomatic` and `halstead` (`volume`, `difficulty`, `effort`) complexity metrics, measuring concepts such as linearly independent paths, storage cost, code understanding effort, coding time, etc. Owing to Frama-C's issues with complex real-world code, for D2A samples we were able to collect `sloc` and `cyclomatic` measures from the Lizard tool [42] (additionally cross-verified with pmccabe's [47] metrics output).

**Signal Aware Recall (SAR):** We use the SAR metric to measure the signal awareness of AI-for-code models, as well as its subsequent enhancement by our approaches. To describe it in context of vulnerability detection, while Recall measures the proportion of the vulnerable samples which the model predicts correctly (i.e. True Positives (TP), rest being False Negatives (FN)), SAR measures for how many of those cases does the model capture the right signals to arrive at its prediction. As proposed by [67], this is derived by (i) subjecting each TP to a Delta-Debugging-style minimization cycle, with the model in the loop, followed by (ii) querying the model for its prediction on each TP sample's 1-minimal version, and finally (iii) checking the 1-minimal for the presence (TP') or absence (FN') of the original program sample's bug. While Recall is defined as TP / (TP + FN), SAR is defined as TP' / (TP' + FN' + FN), where TP = TP' + FN'. By definition, SAR <= Recall, and the expectation is to observe a shortening of this gap in the experiments, with our signal awareness enhancement approaches. We use a relative SAR:Recall ratio (<= 1) to present the results of our experiments, encapsulating how much of the model's Recall is attributable to task-relevant signal learning, and enabling easy comparison across different configurations. As shown in Appendix Table A.1, model performance does not get compromised in our experiments, while we strive towards improving the model's SAR:Recall.

**Baselines:** Previous work for measuring SAR [67] uses a combination of checkers to test bug existence in a TP sample's 1-minimal– utilizing primarily the Infer analyzer [23], with fallback to line-based bug matching for samples with differing Infer verdict and the original bug. As compared to Infer, line-based bug matching is less accurate and counts even partial matches in favor of the model, as shown in Appendix Figure A.1, thereby providing a looser SAR bound than Infer analysis. To show the true impact of SAR improvements with our techniques, which line-based matching would mask, we instead employ the stricter Infer-based matching alone. For correctness, we focus only on the samples where Infer verdict matches the original bug. As a result of using the tighter SAR bound, the model baseline SARs are lower than those reported in previous work. The last row of Table A.1 confirms SAR jumping back up, when incorporating line-based checker as in the original setup. Nevertheless, as can be seen, even in this setting our techniques still record SAR improvements, albeit masked by the checker leniency.

## 5.4 Research Questions

With the setup in place, following are the main research questions we aim to explore with our experiments:

1. What impact does complexity-ranked training have on model learning behavior, and does it improve model signal awareness?

2. What impact does program simplification based augmentation have on model signal awareness, and is it better than generic augmentation?

3. What sort of model learning deduction can be obtained by leveraging the dataset's code complexity distribution, and is it more insightful than usual statistical measures?

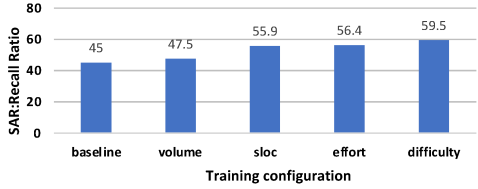

**Figure 9: Model SAR improvements with complexity-ranked training across different complexity metrics. [GNN; s-bAbI ]**

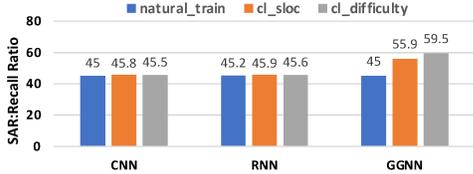

**Figure 10: SAR improvement over baseline (natural_train) with two complexity-ranked training schemes, across models. [s-bAbI ]**

## 6 RESULTS

We first highlight the model signal-awareness improvements achieved by our approaches. Then we show an example of model learning deductions obtained with our data-driven introspection approach.

### 6.1 Code-complexity-ranked Training

Figure 9 shows how an AI-for-code model's signal awareness can be significantly improved by simply presenting it source code samples in the increasing order of their code complexity. Shown is the SAR:Recall ratio for a GNN model on the s-bAbI dataset for different training configurations, including natural random-sampling based training, and four complexity-ranked training schemes based upon sloc, volume, difficulty, and effort complexity metrics respectively. As can be seen, complexity-ranked training can significantly boost the model's signal awareness, with difficulty-ordered training achieving a 32% improvement. However, not all models show such improvements, as shown in Figure 10 comparing CNN, RNN and GNN models across a couple such complexity-ranked training schemes (same results for others; not shown). The fact that other models aren't able to improve, serves to highlight the learning superiority of GNNs, which operate on a more natural graph-based representation of source code than the code-as-image (CNN) and code-as-token-sequence (RNN) counterparts [6, 21, 66].

Figure 11a shows how the model learning changes with complexity-ranked training for the GNN model on the s-bAbI dataset. It shows the validation accuracy curves (i.e. model's interim accuracy on the validation set as it progresses along it's training rounds or 'epochs') for the different training configurations- natural training, and the four complexity-ranked training schemes. As can be seen, natural training quickly reaches quite close to its peak performance, whereas the learning is relatively slow with the complexity-ranked schemes. Although all eventually reach similar looking peaks, the crucial difference is the minima reached by these training configurations. Even though both- natural and complexity-ranked training schemes- reach a 90%+ accuracy (and F1, Recall), the latter is much more task-aware, as shown before in Figure 9's signal aware recall values. This can be attributed to the model learning, albeit slowly, better signals from 'easier' examples first, empowering it to sift-through noise with more complex examples later.

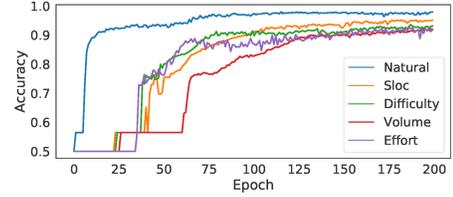

(a) s-bAbI dataset

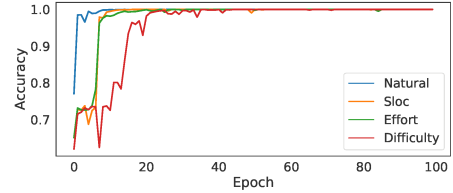

(b) Juliet dataset

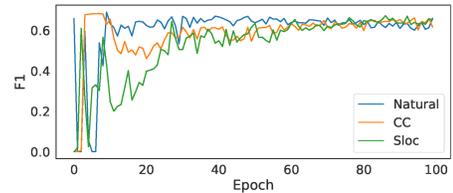

(c) D2A dataset

**Figure 11: Comparing validation performance curves. Complexity-ranked training is slower than natural training, but eventually makes up. F1 curve for D2A for better visualization; accuracy curve noisier but has same trend. (CC = cyclomatic complexity). [GNN]**

This altered training behavior is also seen in the other models and datasets, with Figure 11c showing another such configuration for the GNN model on the real-world D2A dataset (same trend with other configurations; omitted due to space constraints). Although complexity-ranking alters the model's learning/training route, and by itself can offer some assistance to the models, it doesn't seem to always help the model reach a more task-relevant minima. The baseline SAR:Recall, obtained with the natural training scheme, remains unchanged for the Juliet even with complexity-ranked training, while D2A shows only a small 3.5% improvement. But as shall be shown later (Section 6.3), complexity-ranking still has some benefits to offer in those settings.

*Summary: The simplicity of complexity-ranked training, together with the altered model learning, offers some assistance to model signal awareness, although not universally.*

### 6.2 Augmentation via Program Simplification

Our program simplification approach results in the generation of 9x more samples for s-bAbI, 9.6x for Juliet, and 53x for D2A, as a factor of the base dataset size. The varying levels of augmentation are due to the difference in the datasets' sample sizes, which tend to be much bigger for the real-world D2A dataset, as compared to s-bAbI and Juliet (median sloc 36 vs. 9). The bigger the input code sample, the more the number of reduction iterations performed by Delta Debugging, resulting in potentially more valid intermediate samples being generated.

Training the models over these additional (and simplified) samples yields even greater signal awareness improvements, than achieved with complexity-ranked training. This can be seen in Figure 12, showing SAR:Recall values achieved with different levels of augmentation for the GNN model over s-bAbI dataset. The x-axis shows the proportion of samples (in percentage of the base dataset size) randomly selected from the generated set, and added to the base dataset for training, with the leftmost point (x=0) referring to the baseline model performance (same as shown in Figure 9). As can be seen, by introducing just 2% additional simplified program samples into the training mix, the model signal awareness improvement surpasses that achieved with complexity-ranked training. The gains continue with more augmentation, with SAR reaching almost 96% of its attainable max (i.e. Recall) with 50% augmentation, amounting to a 113% improvement over the base model signal awareness. By presenting the model with smaller samples, while still containing the characteristics relevant to the task at hand (e.g. bugs), it seems to be helping the model focus more on task-relevant aspects of code, and less on noise or dataset nuances, while maintaining model performance (Section 5.3:SAR). More concrete insights on how model learning is changing under the covers shall be revealed in Section 6.4, highlighting the potential of our introspection approach.

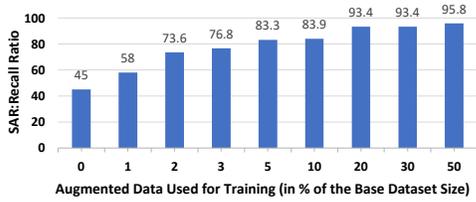

Figure 12: GNN SAR improvements with random sampling over the augmented s-bAbI dataset.

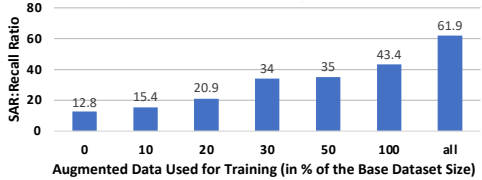

Figure 13: GNN SAR improvements with random sampling over the augmented Juliet dataset.

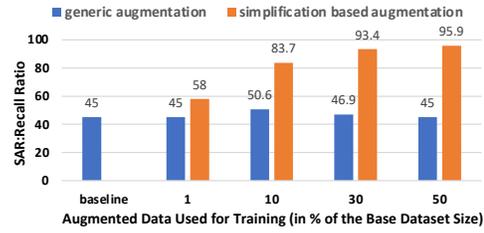

Figure 14: Comparing the generic and program simplification based augmentation approaches on the s-bAbI dataset.

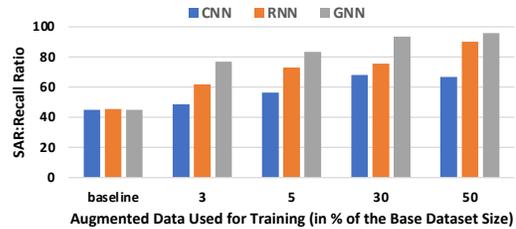

Figure 15: Program simplification based augmentation improves the SAR of CNN, RNN and GNN on the s-bAbI dataset.

The trend is the same for the Juliet dataset, with more augmentation yielding in greater signal awareness improvements, as shown in Figure 13 for a GNN model. The relative gains however are even more extreme, due to the poor baseline model SAR. Model SAR improves dramatically across the augmentation levels, crossing 60% of Recall when *all* generated samples are additionally used for training. The usual model quality metrics (Precision, Recall, etc.) maintain their similar high (90+) values in all augmentation configurations, while SAR increases drastically. This highlights the potential of task-aware learning present in the models, and unlocked by our program simplification based augmentation approach.

D2A records only modest SAR:Recall improvement of 13.3%. Interestingly, the model Recall also gains by 22.1% during augmentation, while the Recalls in the case of s-bAbI and Juliet datasets are stably high. This suggests the base dataset is not sufficiently large for model training. Just with derived simplified examples alone, we are able to guide the model to correctly capture more signals and thus improve both the classic model recall and signal awareness recall. On the other hand, this also points to the diversity of the real-world programs, and the challenge of generating sufficient simplified programs that can help models distinguish various signals from noises in such a diverse code base. Note that the sophisticated methodology behind D2A curation makes it non-trivial to collect more samples and enlarge the training set [82]. It uses differential analysis based on Github commit history, for filtering the false positives from (presumably) a state-of-the-art Infer static analyzer.

These performance gains are not just due to fact that there are extra samples to train upon. This can be seen in Figure 14 which shows the SAR:Recall values obtained with generic augmentation, compared to our approach, for a few representative augmentation levels for the s-bAbI dataset. As can be seen, just adding more samples to the training set does not necessarily increase the model's signal awareness, unlike our simplified program sample augmentation. The fact that the code samples generated by our approach are smaller and potentially simpler that the original samples, is crucial to the model being able to better capture task-relevant signals and sift through noise during training. The results are the same for the Juliet dataset, with generic augmentation not improving upon the baseline SAR at all, irrespective of the augmentation level, very much unlike our augmentation approach. As for D2A, since it is already very limited in size, so there isn't enough hold-out data to test generic augmentation.

Unlike the case with complexity-ranked training, the signal awareness of CNN and RNN models is also given a boost with our augmentation approach. This is shown in Figure 15 showing SAR:Recall values for the three models for a few example augmentation levels (similar trend for other model-dataset configurations; omitted due to space constraints). As can be seen, the GNN model outshines the competitors, as was the case with complexity-ranked training, again highlighting is superior potential for better learning with appropriate guidance.

*Summary: Dataset augmentation with simplified, de-noised program samples assists models in learning task-relevant signals better, while maintaining model performance.*

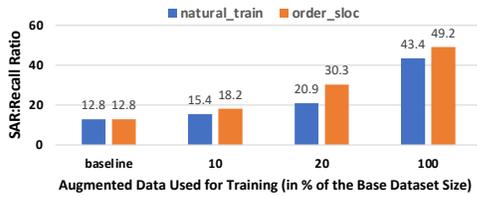

Figure 16: Hybrid training: combining complexity-ranking with program simplification based data augmentation.

## 6.3 Hybrid Training

The two approaches- complexity-ranked training and program-simplification-based augmentation- are complimentary to each other, and can potentially be combined together for different use-cases. These include schemes such as (i) selecting only the more complex samples to be simplified, or (ii) ordering the augmented samples in the order of their code complexity metrics during training, or (iii) training and comparing the model separately on subsets ordered by complexity, for verifying model capacity and quality, amongst others. We experiment with one such hybrid setting to explore the potential for even more gains to be had in the model signal awareness, by performing complexity-ranking atop the training dataset augmented with simplified program samples. Additional signal-awareness boost with such a combination was most evident for the Juliet dataset, as shown in Figure 16 for a few example augmentation configurations. This is in stark contrast with the inability of complexity-ranking by itself to improve model's signal awareness for Juliet (Section 6.1). One hypothesis for this is the expanded metric range which opens up post-simplification, as depicted in Figure 7, improving ranking granularity and thus new sample ordering during training.

## 6.4 Model Learning Introspection

So far, we have shown the potential of different data-driven approaches to improve model signal awareness. Although we have conjectured the possible reasons behind such improvements, we haven't yet probed the model black-box. In this Section, we present the results of our data-driven model introspection approach to analyze model evolution across augmentation iterations.

Our introspection approach deduces model learning behavior by comparing the complexity distributions of the test-set samples, grouped by their prediction accuracy. We use this atop two special classes of samples not affected by augmentations. Specifically, for each model $M_i$ trained under an augmentation setting (i.e. base dataset + $X\%$ simplified samples), subsequent signal-awareness measurement results in it's true positives being divided into SAR-$TP_i$ and SAR-$FN_i$, depending on if $M_i$ captured the real signals or not. Then, the two special classes in focus are: (i) AlwaysTP := $\bigcap_{i=1}^{n}$ SAR-$TP_i$, samples always captured correctly by the model; (ii) AlwaysFN := $\bigcap_{i=1}^{n}$ SAR-$FN_i$ that are consistently mispredicted. The intersection of SAR-$TP_i$ or SAR-$FN_i$ allows us to focus on samples that are not affected by the augmentations, and thus examine the characteristics of both- the straightforward and challenging samples- for a model architecture.

Figure 17(a) compares the difficulty complexity metrics distribution of the AlwaysTP samples for the s-bAbI dataset, versus the AlwaysFN group. It provides an interesting insight into the model learning behavior from the point of view of the difficulty of the program to write or understand. Across all augmentation iterations, the model is more easily able to correctly capture less difficult code samples (difficulty = {12,14}), while consistently mis-predicted samples tend to be harder (difficulty = {16,17}). This does not mean that the latter category is never predicted correctly by the model. The model eventually learns to be able to predict them with sufficient augmentation, but the ones that still remain mispredicted possess the higher metric scores.

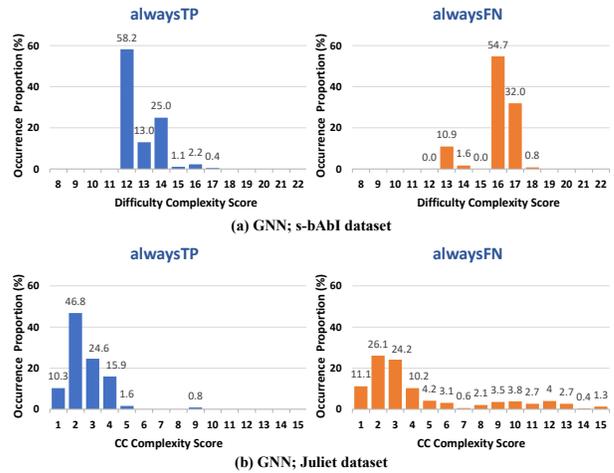

Figure 17: Comparing characteristics of samples consistently (mis)predicted by model across augmentations

Figure 17(b) shows the corresponding behavior for the model augmentation on the Juliet dataset, this time taking the example of the cyclomatic complexity (cc) metric (# independent paths in the program). The model is able to somewhat capture low complexity samples (cc < 5) as opposed to the more complex ones (cc > 10), where it fails consistently, irrespective of augmentation assistance. The separation between the distributions for AlwaysTP and AlwaysFN is not as clear as for s-bAbI, since model learning isn't as good for Juliet to begin with, with almost half of the test-set samples are consistently mispredicted for Juliet despite augmentation.

Using other metrics such as ifs, loops etc. yields more fine grained insight into model learning. For ex., for s-bAbI, across augmentation iterations, the model is more easily able to correctly capture code samples containing no loops (77%), while consistently mis-predicted samples tend to have a loop in them (88%).

This is one kind of insight that can be derived using our data-driven model introspection approach. Beyond focusing on these special classes, we present one more use-case of our model introspection approach- tracing how model understanding of source code improves across augmentation iterations. As before, we use the comparison between the code complexity distribution of SAR-TP and SAR-FN samples to analyze model evolution from the dataset's perspective. Figure 18 presents this comparison using sloc distribution across s-bAbI augmentation iterations. Similar to the trend uncovered in Figure 17(a), the first insight in this case is regarding the baseline model (leftmost, no augmentation) facing trouble understanding bigger code samples (sloc = {12,13}). This weakness improves as augmentation increases, as can be seen with the rising

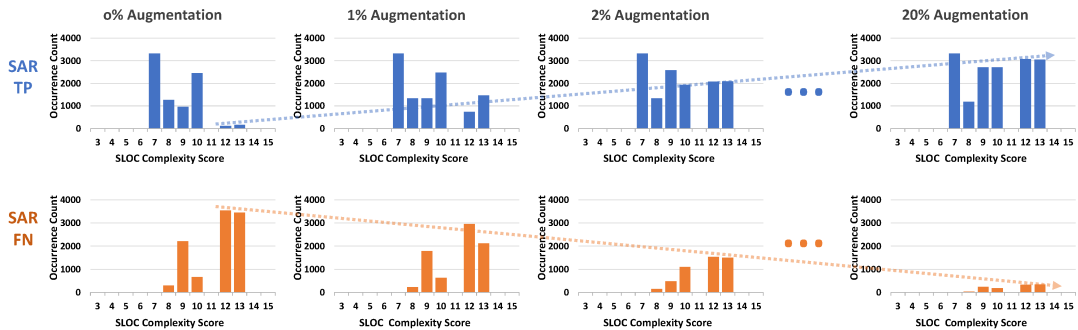

Figure 18: Insight: "More augmentation is helping model better understand bigger code samples"- via complexity distribution comparison between SAR-TP/FN groups. "X% Augmentation" = base dataset + X% augmented samples (in % of base dataset size)

SAR-TP 'skyline' (i.e. sample occurrence counts), and correspondingly the falling SAR-FN skyline, most evident for sloc = {12,13} samples. This leads to an intriguing insight about augmentation helping the model better understand bigger code samples. This is especially interesting because the model learning was generic- neither the augmentation scheme nor the model were aware of any notion of code complexity. It is only after the fact that we analyze the model's prediction performance from the perspective of the test set's code complexity, that we uncover these findings. Different insights can be derived by changing the complexity metrics employed for such data-driven model introspection.

*Summary: Model introspection from the dataset perspective yields code-centric, developer-friendly insights into model learning behavior, beyond the usual generic measures of model quality.*

## 7 RELATED WORK

The most relevant work using Delta Debugging (DD) and its variant methods is software failure diagnosis and isolation [31, 49, 78, 79]. The main advantage of DD is that it can significantly reduce the number of tests needed to locate the problem. While DD has recently been employed to expose the low signal-awareness problem in AI-for-code models [67], as well as model explanation [57], we use it to assist the models to improve their signal awareness.

We do this by simplifying the training set samples using DD, and augmenting them back into the training mix. Augmentation methods are popular in AI in general, including domain-specific approaches such as image transformation based on geometry, color, overlap, etc.,[63], text transformations [15], model-specific approaches such as adversarial training [45], and data-driven approaches such as SMOTE [17], formal and empirical augmentation [36, 50, 77], amongst others. Complementary to these approaches, our augmentation approach is focused specifically towards improving the signal awareness of source code models. A key difference is that general AI approaches usually assume the input under augmentation would keep the original labels since it is extremely hard to check for images and texts without huge manual effort, while this assumption may not always be true. Our approach utilize the benefits of working in well-defined source code space, therefore we can assure the validity and correctness of our code augmentation.

Apart from augmentation, we improve signal awareness of AI-for-code models by incorporating more code-specific knowledge into their training. We do this by combining the SE-concept of code complexity with the AI approach of curriculum learning [11, 28, 32]. While CL has previously been applied using general complexity measures of images and texts to rank training samples in the vision and natural language domains, we use code-specific complexity measures to tailor the models towards source code understanding.

Finally, we also use the notion of code complexity to introspect model learning. It is hard to understand what deep learning models learn, to address which methods have been developed for AI in general. This includes probing the model's gradient [61, 64, 65, 83] to highlight input regions most influencing the model's prediction, or fitting interpretable surrogate models to approximate the deep learning model's behavior [30, 44, 58], and then using the surrogate to derive the feature importance ranking for the input. Explanation methods have also been created for graphical neural networks, attributing importance to a graphs nodes and edges by using attention mechanisms [72, 73], or via maximizing mutual information between inputs and outputs [76]. Our approach is complementary to these approaches as it treats models as black boxes, deducing model learning from the dataset's perspective, and offers more code-centric insights. Please see Section 1 for a contrast with AI-for-code specific model explanation approaches.

## 8 CONCLUSION

In this paper we presented multiple SE-assisted data-driven techniques to assist and introspect AI-for-code model learning. We incorporated the notion of complexity of source code into model training to enhance model signal-awareness, achieving significant improvements with our complexity-ranked training and program-simplification-based augmentation approaches. We carried the notion of complexity into model introspection, and presented code-centric insights into black-box model learning. Moving forward, we look to explore active learning approaches, helping the model with targeted SE-assistance, e.g. picking specific samples for augmentation where the model is facing difficulty learning.

## A APPENDIX

```
void CWE121_memcpy_81_bad(int * data) {      void CWE121_memcpy_81_bad(int *data) {
  int source[10] = {0};                        int source[10] = {};
  memcpy(data, source, 10*sizeof(int));        (data, source, 10 * sizeof(int));
  printIntLine(data[0]);                       printIntLine(data[0]);
}                                            }
        (a) Original Code Snippet                   (b) Reduced 1-minimal Snippet
      Model prediction :  Buggy                  Model prediction :  Buggy
      Ground Truth     :  Buggy                  Ground Truth     :  Non-buggy
```

**Figure A.1:** A buffer overflow example showing why a looser SAR bound is measured by line-based bug matching. Even though model incorrectly considers the 1-minimal as buggy, a line-based checker will count this partial match in favor of model capturing real signals, since buggy line exists in 1-minimal.

**Table A.1:** Using {Juliet + GNN + augmentation} example to show: (i) model perf is maintained (rows 1 and 2) as SAR improves (rows 3 and 4) with our augmentation approach, and (ii) signal awareness improvements still achieved with our approach, even when using a looser SAR bound (SAR' in row 5).

|              | base + 0% aug | +10% | +20% | +30% | +50% | +100% | +all |
|--------------|---------------|------|------|------|------|-------|------|
| F1           | 99.9          | 99.9 | 99.9 | 99.9 | 99.9 | 99.9  | 99.9 |
| Recall       | 99.9          | 99.9 | 99.9 | 99.7 | 99.6 | 99.7  | 99.8 |
| SAR          | 12.8          | 15.4 | 20.9 | 33.9 | 34.9 | 43.3  | 61.8 |
| SAR:Recall % | 12.8          | 15.4 | 20.9 | 34.0 | 35.0 | 43.4  | 61.9 |
| SAR'         | 48.4          | 51.1 | 54.7 | 61.1 | 61.2 | 65.3  | 74.8 |